\documentclass[sigconf]{acmart}

\copyrightyear{2020}
\acmYear{2020}

\setcopyright{rightsretained}

\usepackage{url}

\usepackage{amssymb}
\usepackage{amsmath}
\usepackage{amsthm}
\usepackage[ruled]{algorithm2e}

\usepackage{comment}
\usepackage{multirow}

\renewcommand{\textrightarrow}{$\rightarrow$}

\begin{document}
\fancyhead{} 

\title{MAMO: Memory-Augmented Meta-Optimization for Cold-start Recommendation}


\author{Manqing Dong}
\affiliation{
\institution{University of New South Wales}
}
\email{dongmanqing@gmail.com}

\author{Feng Yuan}
\affiliation{
\institution{University of New South Wales}
\city{Sydney}
\country{Australia}
}

\author{Lina Yao}
\affiliation{
\institution{University of New South Wales}
}
\email{lina.yao@unsw.edu.au}

\author{Xiwei Xu}
\affiliation{
\institution{Data 61, CSIRO}
\city{Sydney}
\country{Australia}
}

\author{Liming Zhu}
\affiliation{
\institution{Data 61, CSIRO}
\city{Sydney}
\country{Australia}
}


\begin{abstract}
A common challenge for most current recommender systems is the cold-start problem. Due to the lack of user-item interactions, the fine-tuned recommender systems are unable to handle situations with new users or new items.
Recently, some works introduce the meta-optimization idea into the recommendation scenarios, i.e. predicting the user preference by only a few of past interacted items.
The core idea is learning a global sharing initialization parameter for all users and then learning the local parameters for each user separately. 
However, most meta-learning based recommendation approaches adopt model-agnostic meta-learning for parameter initialization, where the global sharing parameter may lead the model into local optima for some users.
In this paper, we design two memory matrices that can store task-specific memories and feature-specific memories. Specifically, the feature-specific memories are used to guide the model with personalized parameter initialization, while the task-specific memories are used to guide the model fast predicting the user preference. And we adopt a meta-optimization approach for optimizing the proposed method.
We test the model on two widely used recommendation datasets and consider four cold-start situations. The experimental results show the effectiveness of the proposed methods.
\end{abstract}



\maketitle

\section{Introduction}
Personalized recommender systems are playing more and more important roles in web and mobile applications.
Despite the success of traditional matrix-factorization based recommendation methods~\cite{bokde2015matrix} or the most current deep-learning based techniques~\cite{zhang2019deep}, a common challenge for most recommendation methods is the cold-start problem~\cite{wei2017collaborative}. Because of the lack of user-item interactions, the recommendation approaches that utilize such interactions are unable to handle the situations where new users or items exist, corresponding to the user cold-start and item cold-start problems.

Traditional way to solve the cold start problem is leveraging auxiliary information into the recommender systems, e.g.,content-based recommender systems~\cite{roy2016latent,wei2016collaborative} and cross-domain recommenders~\cite{li2018cross,wang2019cdlfm}. For example, to address the item cold-start problem, \cite{wei2016collaborative} proposes a hybrid model in which item features are learned from the descriptions of items via a stacked denoising autoencoder and further combined into a collaborative filtering model timeSVD++. In \cite{wang2019cdlfm}, the authors propose a cross-domain latent feature mapping model, where the neighborhood-based cross-domain latent feature mapping method is applied to learn a feature mapping function for each cold-start user. However, the major limitation of such approaches is that the learned model may recommend same items for users with similar content thereby neglect the personal interests.

Inspired by recent works in few-shot learning \cite{wang2019few}, 
meta-learning \cite{vanschoren2018meta} has been introduced into the recommender systems to solve the cold-start problem, i.e. predicting a user's preferences by only a few past interacted items. Most of the current meta-learning based recommender systems~\cite{chen2018federated,lee2019melu,zhao2019learning} adopt optimization-based algorithms such as model-agnostic meta-learning (MAML)~\cite{finn2017model}, for their promising performance in learning configuration initialization for new tasks.
Generally, the recommendation for a user is regarded as a learning task.
The core idea is learning a global parameter to initialize the parameter of personalized recommender models. 
The personalized parameter will be locally updated to learn a specific user's preference, and the global parameter will be updated by minimizing the loss over the training tasks among the users. Then, the learned global parameter is used to guide the model settings for new users. 
For example, \cite{lee2019melu} uses several fully connected neural networks as recommender models. They first learn user and item embedding from the user and item profiles, and then feed them into the recommender model to get predictions. They define two groups of global parameters: parameters in learning embeddings and parameters in recommender models. They locally update the recommender model for personalized recommendation and globally update the two groups of parameters for the initialization of new users.
\cite{du2019sequential} also constructs a deep neural net as the recommender model. They define a global parameter for initializing the recommender model, and then locally updates the model parameter by introducing update and stop controllers.

The aforementioned approaches have been proven to be promising with regards to applying meta-optimization in the cold-start scenarios and presenting competitive performance in warm-start scenarios. However, they have the following limitations. Most of them are build upon MAML algorithm and its variants, which are powerful to cope with data sparsity, while having a variety of issues such as instability, slow convergence, and weak generalization. In particular, they often suffer from gradient degradation ending up with a local optima when handling users who show different gradient descent directions comparing with the majority of users in the training set. 
To address this issue, we propose Memory-Augmented Meta-Optimization (MAMO) for cold-start recommendation. In details, \textbf{1)} for solving the problem of local optima, we design a feature-specific memory to provide a personalized bias term when initializing the model parameters. Specifically, the feature-specific memory includes two memory matrices: one stores the user profile memory to provide retrievable attention values, and the other caches the fast gradients of the previous training sets to be read according to the retrievable attention values.
\textbf{2)} We further design a task-specific memory cube, i.e. user preference memory, to learn to capture the shared potential user preference commonality on different items. It is used as fast weights for the recommender model to alleviate a need for storing copies of neural activity patterns.
\textbf{3)} The extensive experiments on two widely used datasets show MAMO performs favorably against the state-of-the-arts. The code is publicly available for reproduction\footnote{\url{https://github.com/dongmanqing/Code-for-MAMO}}.

The rest of the paper is organized as follows. We review proposed approach in Section \ref{sec:section2}; Section \ref{sec:sec_3} presents the settings, experimental results and analysis of the experiments; we review the related work in Section \ref{sec:sec_4}, followed by the conclusion in Section \ref{sec:sec_5}.
\section{Proposed Approach}
\label{sec:section2}
\subsection{Overview}
\subsubsection{Problem Definition}
We consider the recommendation for a user as one task. Given a user $u$ with profile $p_{u}$ and rated items $\mathcal{I}_{u}^{S}$, where each item $i$ is associated with a description file $p_{i}$ and corresponding ratings $y_{u,i}^{S}$, our goal is predicting the rating $y_{u,i}^{Q}$ by user $u$ for a new item $i\in \mathcal{I}_{u}^{Q}$. Here, $S$ stands for the support set, and $Q$ stands for the query set. 

\subsubsection{Motivation}
Most existing meta learning based recommendation techniques attempt to learn a meta global parameter $\phi$ to guide the model initialization $\mathcal{R}_{\theta}$, i.e. $\theta \leftarrow \phi$, where $\phi$ is learned across all the users in training set. It provides a uniform parameter initialization to govern the trained recommendation model to predict on new users. 
The global parameter $\phi$ works uniformly across all users and thus may be inadequate to discern the intrinsic discrepancies among a variety of user patterns, resulting in poor generalization. Also, the model may tend to be stuck on a local optimum and encounter gradient degradation. 
Instead of learning a single initialization of the model parameters, we propose an adaptive meta learning recommendation model, Memory-Augmented Meta-Optimization (MAMO), to improve the stability, generalization and computational cost of model by learning a multi-level personalized model parameters. 

\subsubsection{Model Structure}
The proposed model includes two parts: the recommender model for predicting the user preference and the memory-augmented meta-optimization learner for initializing recommender model parameters.
The recommender model $\mathcal{R}_{\Theta}(p_{u},p_{i})$ predicts the recommendation scores, where the model parameters $\Theta=\{\theta_{u}, \theta_{i}, \theta_{r}\}$ will be locally updated for single users to provide personalized recommendation.
The meta-learner, which includes global parameters $\Phi=\{\phi_{u}, \phi_{i}, \phi_{r}\}$ and memories $M=\{M_{U}, M_{P}, M_{U,I}\}$, will provide personalized initialization for the recommender model parameters $\Theta$, and will be globally updated during the training process for users $u\in U^{train}$.
The learned meta-learner is further used to initialize the recommender model for new users $u\in U^{test}$.

\begin{figure}[t]
    \centering
    \includegraphics[width=\linewidth]{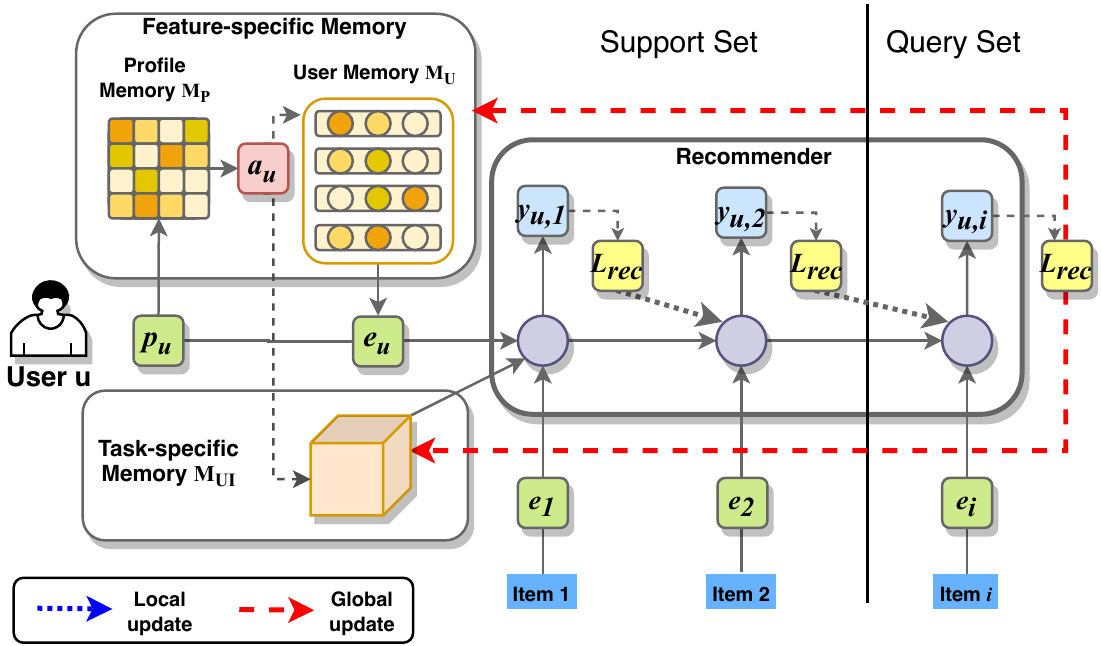}
    \caption{The training phase of MAMO}
    \label{fig:structure_training}
\end{figure}

\subsection{Recommender Model}
Similar to many previous works, we assume that the user preference is coming from a complex combination of his/her personal information such as user profiles, user rating records, and the item profiles.

\subsubsection{Embedding: $p_{u}\rightarrow e_{u}$, $p_{i}\rightarrow e_{i}$}
We use user profile $p_{u}$ to represent the initial user preference for the following considerations. First, user profile normally includes general information such as age groups and locations. In a cold-start scenario, where there are only limited user-item interaction records, this will provide potential user preference for the recommenders.
Second, traditional one-hot user representation (by a unique id) is strongly relied on collaborative filtering techniques. The fine-tuned model is hard to adapt to new users. Similarly, we use $p_{i}$ to denote the item profile for learning the item embedding.

To learn the user and item embedding vectors $e_{u}\in \mathbb{R}^{d_{e}}$ and $e_{i}\in \mathbb{R}^{d_{u}}$ from $p_{u}\in \mathbb{R}^{d_{u}}$ and $p_{i}\in \mathbb{R}^{d_{i}}$, the simplest way is constructing a multi-layer fully connected neural network, i.e.
\begin{equation}
\label{eq:embedding}    e_{u} = f_{\theta_{u}}(p_{u}), e_{i} = f_{\theta_{i}}(p_{i})
\end{equation}
where $d_{e}$ is the embedding size; $f$ denotes the fully connected layers; $d_{u}$ and $d_{i}$ stand for the dimension of the user profile vector and item profile vector; $\theta_{u}$ and $\theta_{i}$ are the parameters of the fully connected layers for learning user and item embedding, respectively. We will omit similar notations in the rest of the paper.

\subsubsection{Recommendation: $(e_{u}, e_{i})\rightarrow y_{u,i}$}
\label{sec:recommendation}
Given the user embedding $e_{u}$, and a list of rated item embeddings $e_{i}$ for $i\in \mathcal{I}^{Q}_{u}$, we get the prediction of preference score $y_{u,i}$ for each item by:
\begin{equation}
\label{eq:rec} \hat{y_{u,i}} = f_{\theta_{r}}(e_{u}, e_{i}) = FC_{\theta_{r}}(M_{u,I} \cdot [e_{u}, e_{i}])
\end{equation}
where $[e_{u},e_{i}]$ is the concatenation of the user embedding and the item embedding, and $FC$ denotes the fully connected layers. $M_{u,I} \in \mathbb{R}^{d_{e}\times 2d_{e}}$ is a matrix that includes the fast weights of the recommender model for user $u$, which is extracted from the task-specific memory $M_{U,I}$ according to the user profile $p_{u}$, i.e. $M_{u,I} \leftarrow (p_{u}, M_{U,I})$. 
The task-specific memory $M_{u,I}$ for user $u$, will be locally updated (i.e. during the learning on support set of user $u$) for personalized recommendation. We will introduce more details about the task-specific memory in the further parts.

\subsection{Memory-Augmented Meta-Optimization}
\subsubsection{Feature-specific Memory: $(p_{u}, M_{U}, M_{P}) \rightarrow b_{u}$.}
Recall that the parameters used for extracting user and item embedding are $\theta_{u}$ and $\theta_{i}$, traditional meta optimization approach will learn the global parameters $\phi_{u}$ and $\phi_{i}$ for initialization, i.e. $\theta_{u} \leftarrow \phi_{u}$, $\theta_{i} \leftarrow \phi_{i}$. 
Here for addressing the single initialization problem, we introduce feature-specific memories, i.e. the user embedding memory $M_{U}$ and the profile memory $M_{P}$, to help the model initialize a personalized parameter $\theta_{u}$. 
The profile memory $M_{P}$ stores information relevant to user profiles $p_{u}$ to provide the retrieval attention values $a_{u}$. The retrieval attention values are applied for extracting information from user embedding memory $M_{U}$, where each row of $M_{U}$ keeps the according fast gradients (or bias terms). 
Together, the two memory matrices will help to generate a personalized bias term $b_{u}$ when initializing $\theta_{u}$, i.e. $\theta_{u} \leftarrow \phi_{u} - \tau b_{u}$. Here the bias term $b_{u}$ can be regarded as a personalized initialization parameter for guiding the global parameter $\phi_{u}$ to fast adapt to the case of user $u$; the $\tau$ is a hyper-parameter for controlling how much bias term is considered when initializing $\theta_{u}$.

In details, given a user profile $p_{u}$, the profile memory $M_{P}\in \mathbb{R}^{K \times d_{u}}$ will calculate the attention values $a_{u}\in \mathbb{R}^{K}$ by
\begin{equation}
\label{eq:mu_attention}    a_{u} = attention(p_{u}, M_{P})
\end{equation}
where $attention$ function calculates the cosine similarity between the user profile and the user profile memory, and then be normalized by softmax function. The dimension $K$ denotes the number of user preference types, which is a predefined number before the training process.
We obtain the personalized bias term $b_{u}$ by
\begin{equation}
\label{eq:mu_bu}    b_{u} = a_{u}^{\top} M_{U}
\end{equation}
where $M_{U}\in \mathbb{R}^{K \times \{d_{\theta_{u}}\}}$ stores the memory of the fast gradients. Notice that the user embedding model may comprise more than one neural layer and more than one parameter, which means $M_{U}$ is not a numerical matrix but stores all the fast gradients with the same shape as the parameters in the user embedding layers, so here $d_{\theta_{u}}$ denotes the dimension of the parameters in user embedding layers.

Before the training process, the two memory matrices are randomly initialized, and will be updated during the training process. 
Specifically, profile memory will be updated by:
\begin{equation}
\label{eq:profile_global_update}  M_{P} = \alpha \cdot (a_{u}p_{u}^{\top}) + (1-\alpha) M_{P}
\end{equation}
where $(a_{u}p_{u}^{\top})$ is the cross product of $a_{u}$ and $p_{u}$, $\alpha$ is a hyper-parameter to control how much new profile information is added. Here we add an attention mask $a_{u}$ when add the new information so that the new profile information will be attentively added to the memory matrix.
Similarly, the parameter memory $M_{U}$ will be updated by
\begin{equation}
\label{eq:embedding_global_update} M_{U} = \beta \cdot (a_{u} \nabla_{\theta_{u}}(\mathcal{L}(\hat{y_{u,i}}, y_{u,i}))) + (1-\beta) M_{U}
\end{equation}
where $\mathcal{L}(\hat{y_{u,i}}, y_{u,i})$ denotes the training loss, and $\beta$ is the hyper-parameter to control how much new information is kept. 

\subsubsection{Task-specific Memory: $(a_{u}, M_{U,I}) \rightarrow M_{u,I}$}
The user preference matrix $M_{u,I}\in \mathbb{R}^{d_{e}\times 2d_{d}}$ serves as fast weights or a transform matrix for the recommender model from the user and item embedding (see equation (\ref{eq:rec})) that is extracted from the memory cube $M_{U,I}\in \mathbb{R}^{K \times d_{e}\times 2d_{e}}$, where $K$ is the same notation for the dimension of feature-specific memory that denotes the number of user preference type. Similar to the feature-specific memory, which follows the idea of Neural Turing Machine (NTM)\cite{graves2014neural}, the memory cube $M_{U,I}$ have a read head to retrieve the memory and a write head to update the memory.
Similarly, we attentively retrieve the preference matrix $M_{u,I}$ from $M_{U,I}$ by:
\begin{equation}
\label{eq:mui} M_{u,I} = a_{u}^{\top}\cdot M_{U,I}
\end{equation}
where $a_{u}\in \mathbb{R}^{K}$ is learned from equation \ref{eq:mu_attention}. 
The write head will write the updated personal preference memory matrix $M_{u,I}$ to the $M_{U,I}$ after the learning on the support set. 
\begin{equation}
\label{eq:preference_global_update}    M_{U,I} = \gamma\cdot (a_{u}\otimes M_{u,I}) + (1-\gamma) M_{U,I}
\end{equation}
where $\otimes$ denotes the tensor product, $\gamma$ is a hyper-parameter to control how much new preference information is added. 

\begin{algorithm}[t]
\caption{Training process of MAMO}
\label{alg:training_process}
\LinesNumbered
\KwIn{Training user set $U^{train}$; User profile $\{p_{u}|u\in U^{train}\}$; Item profile $\{p_{i}|i\in (I^{S}_{u}, I^{Q}_{u})\}$; User ratings $\{y_{u,i}|u\in U^{train}, i\in (I^{S}_{u},I^{Q}_{u}) \}$; Hyper-parameters $\alpha$, $\beta$, $\gamma$, $\tau$, $\rho$, $\lambda$}
\KwOut{Meta parameters $\phi_{u}$, $\phi_{i}$, $\phi_{r}$, $M_{P}$, $M_{U}$, $M_{U,I}$}
Randomly initialize the meta parameters $\phi_{u}$, $\phi_{i}$, $\phi_{r}$\;
Randomly initialize the memories $M_{P}$, $M_{U}$, $M_{U,I}$\;
\While{Not Done}{
    \For{$u\in U^{train}$}{
        Calculate bias term $b_{u} \leftarrow (p_{u}, M_{U}, M_{P})$ by Eq.~(\ref{eq:mu_attention}-\ref{eq:mu_bu})\;
        Initialize the local parameters $\theta_{i}$, $\theta_{u}$, $\theta_{r}$ by Eq.~(\ref{eq:local_initialize1}-\ref{eq:local_initialize2})\;
        Initialize the preference memory $M_{u,I}\leftarrow (a_{u}, M_{U,I})$ by Eq.~(\ref{eq:mui})\;
        \For{$i \in I^{S}_{u}$}{
            Get user and item embedding $e_{u}$ and $e_{i}$ by Eq.~(\ref{eq:embedding})\;
            Get prediction of $\hat{y_{u,i}}$ by Eq.~(\ref{eq:rec})\;
            Local update $M_{u,I}$\;
            Local update $\theta_{u}$, $\theta_{i}$, $\theta_{r}$ by:
            $\theta_{*} \leftarrow \theta_{*} - \rho \cdot \nabla_{\theta_{*}}\mathcal{L}(y_{u,i}, \hat{y_{u,i}})$\;
        }
    }
    Update feature-specific memory $M_{P}$, $M_{U}$ by Eq.~(\ref{eq:profile_global_update}-\ref{eq:embedding_global_update})\;
    Update task-specific memory $M_{U,I}$ by Eq.~(\ref{eq:preference_global_update})\;
    Update global parameters $\phi_{u}$, $\phi_{i}$, $\phi_{r}$ by:
    $\phi_{*} \leftarrow \phi_{*} - \lambda \Sigma_{u\in U^{train}}\Sigma_{i\in \mathcal{I}_{u}^{Q}} \nabla \mathcal{L}(\mathcal{R}_{\hat{\theta_{*}}})$\;
}
\end{algorithm}

\subsubsection{Local Update}
Traditionally, the parameters of a neural network are initialized by randomly sampling from a statistical distribution. Given sufficient training data, the randomly initialized parameters can usually converge to a good local optimum but may take a long time \cite{du2019sequential}.
In the cold-start scenario, random initialization combined with limited training data can lead to serious over-fitting, which makes the trained recommender insufficient to generalize well.
Inspired by recent works of meta-training\cite{lee2019melu}, we initialize the recommender parameters from the global initial values.

At the beginning of the training process, we randomly initialize the global parameters, i.e. $\phi_{u}$, $\phi_{i}$, $\phi_{r}$, $M_{U}$, $M_{P}$, $M_{U,I}$.
For each user $u\in U^{train}$, we have support set and query set. During the local learning phase (i.e. learning on the support set), we initialize the local recommender parameters by:
\begin{align}
\label{eq:local_initialize1}    \theta_{u} &\leftarrow \phi_{u} - \tau b_{u}\\
\label{eq:local_initialize2}    \theta_{i} &\leftarrow \phi_{i}, \theta_{r} \leftarrow \phi_{r}
\end{align}
where $b_{u}$ is obtained via equation~(\ref{eq:mu_attention}-\ref{eq:mu_bu}). Then we obtain the task-specific memory matrix $M_{u,I}$ by equation~(\ref{eq:mui}).
The prediction of ratings for items $i\in \mathcal{I}^{S}_{u}$ is based on equation~(\ref{eq:rec}). 
The optimization goal in local training is to minimize the loss of the recommendation for a single user, i.e. updating the local parameters by minimizing the prediction loss $\mathcal{L}(y_{u,i}, \hat{y_{u,i}})$. Thus, the local parameters will be updated by:
\begin{equation}
\label{eq:local_update}    \theta_{*} \leftarrow \theta_{*} - \rho \cdot \nabla_{\theta_{*}}\mathcal{L}(y_{u,i}, \hat{y_{u,i}})
\end{equation}
where $*$ could be either $i$, $u$, or $r$; $\rho$ is the learning rate for updating the local parameters. The preference matrix $M_{u,I}$ will also be locally updated via back-propagation. 

\subsubsection{Global Update}
The aim of the meta optimization is to minimize the expected loss on the local query set $i\in \mathcal{I}_{u}^{Q}$ for $u\in U^{train}$. 
Here, the parameters of the meta-learner include: shared initial parameters $\phi_{u}$, $\phi_{i}$ and $\phi_{r}$; feature-specific memories, i.e. profile memory $M_{P}$ and user embedding memory $M_{U}$; and the task-specific memory $M_{U,I}$.
The gradients related to the meta-testing loss, which we call meta-gradient, can be computed via back-propagation. The meta-gradient may involve higher-order derivatives, which are expensive to compute when the depth of the neural nets is deep. Therefore, MAML\cite{santoro2016meta} takes one-step gradient descent for meta-optimization. 
We take similar ideas, where after the local training on the support set, we update the global parameters according to the loss on query sets. 

Suppose the recommender model is denoted as $\mathcal{R}_{\theta}$ combined with a task-specific memory $M_{u,I}$ for user $u$, where $\theta=\{\theta_{u}, \theta_{i}, \theta_{r}\}$. 
After the local training on support set, we get the model $\mathcal{R}_{\hat{\theta}}$ with updated parameters $\hat{\theta}$.  
Our goal is minimizing the training loss for users $u\in U^{train}$ on query sets for $i\in \mathcal{I}^{Q}_{u}$.
Then, the global parameters are updated by
\begin{equation}
    \phi_{*} \leftarrow \phi_{*} - \lambda \Sigma_{u\in U^{train}}\Sigma_{i\in \mathcal{I}_{u}^{Q}} \nabla \mathcal{L}(\mathcal{R}_{\hat{\theta_{*}}})
\end{equation}
Meanwhile, the feature-specific memories $M_{P}$ and $M_{U}$ will be updated by equation~(\ref{eq:profile_global_update}) and ~(\ref{eq:embedding_global_update}); the task-specific memory $M_{U,I}$ will be updated by equation~(\ref{eq:preference_global_update}).
The pseudo code of the training process is listed in algorithm~\ref{alg:training_process}.

\section{Experiments}
\label{sec:sec_3}
\subsection{Datasets}
\subsubsection{Datasets.}
We use two widely used public available datasets for evaluation: MovieLens 1M\footnote{\url{https://grouplens.org/datasets/movielens/}} and Book-crossing\footnote{\url{http://www2.informatik.uni-freiburg.de/~cziegler/BX/}}, which have both user information and item information.
MovieLens 1M includes around 1 million ratings from about 6 thousand users for over 3 thousand movies, and the ratings range from 1 to 5. 
For the Book-crossing dataset, we filter the records that users or items do not have relevant profiles. 
The processed dataset includes about 600 thousand ratings from around 50 thousand users for over 50 thousand books, the ratings are between 1 and 10. See appendix \ref{sec:dataset_details} for more details.

\subsubsection{Data preprocessing}
We use different strategies when learning the feature representation. For features have single categorical value, such as location and occupation, each feature is denoted by index and is represented with a randomly initialized embedding vector; we use similar approaches to process the numerical features such as the publication year and age group. 
For features that may have multiple categories, such as movie genres and directors, we use one-hot representation and then transform them into vectors that have same dimension with other feature vectors. 

\subsubsection{Training and testing dataset.}
Each user is regarded as a sample in the dataset. We randomly separate the users into training and testing users with ratio 80:20. The history records for a single user are further divided into support set and query set. 
We trim the number of rating records for each user as with 20 records to force the model to learn from few samples. 
In default settings, we consider the first 15 items for a user as the support set and the others as the query set. 
A detail is that we sort these items according to the user review time. By doing so, the rated items in query set are logically regarded as 'new items' for a user. 
We will further discuss whether the ratio of training dataset or the number of cases in the support set will affect the model performance. 

\subsubsection{Cold-start scenarios.}
\label{sec:cold_start}
We further consider the recommendation performance on four scenarios: 1) existing users for existing items (W-W) ; 2) existing users for cold items (W-C); 3) cold users for existing items (C-W); and 4) cold users for cold items (C-C).
For the MovieLens dataset, we classify the users into warm or cold users according to their first comment time. The first comment time in the MovieLens dataset ranges from 2000-04-26 to 2003-03-21. We found most users (about 90\%) provide their first comments before 2000-12-03. By dividing the users based on this time, we have 5,400 warm users and 640 cold users. As for the items, we regard items with less than 10 ratings as cold items, where we get 1,683 warm items and 1,645 cold items.
For the Book-crossing dataset, where the rating time is not provided, we assume the order of the user ids is consistent with the order of the user registration time. Thus, we assume the first 90 percent of the users are warm users and the others are cold users. Similarly, we regard items with more than 5 ratings as warm items, from which we get 628 warm items and the others as cold items.

\begin{figure*}[t]
\centering
\begin{minipage}[t]{4cm}
\includegraphics[width=0.9\textwidth]{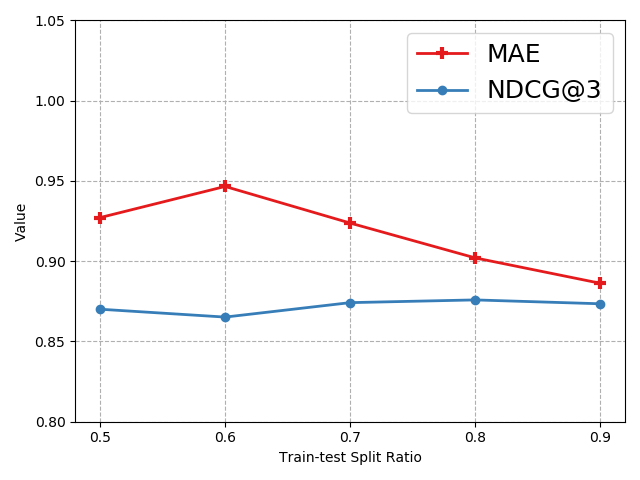}
\centering{(a)}
\end{minipage}
\begin{minipage}[t]{4cm}
\includegraphics[width=0.9\textwidth]{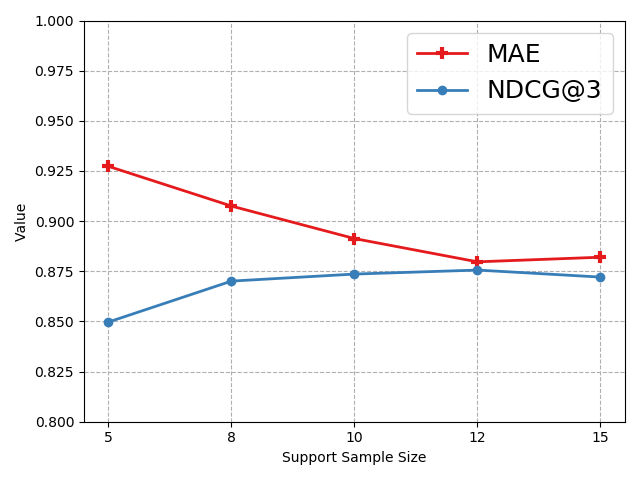}
\centering{(b)}
\end{minipage}
\begin{minipage}[t]{4cm}
\includegraphics[width=0.9\textwidth]{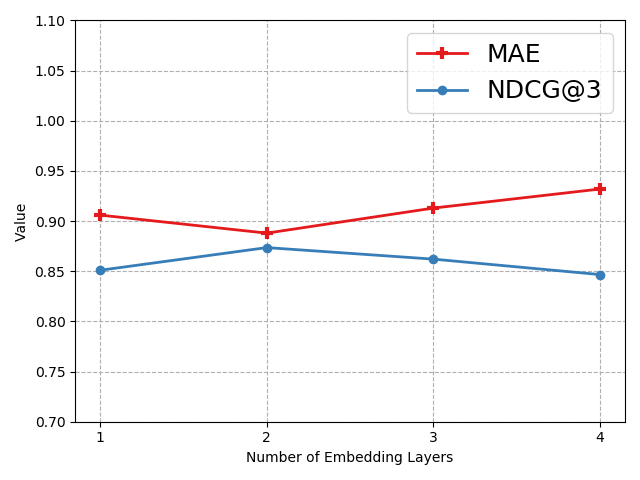}
\centering{(c)}
\end{minipage}
\begin{minipage}[t]{4cm}
\includegraphics[width=0.9\textwidth]{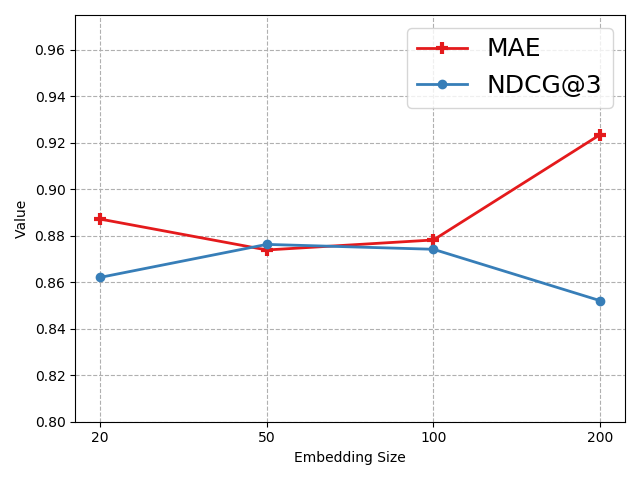}
\centering{(d)}
\end{minipage}
\begin{minipage}[t]{4cm}
\includegraphics[width=0.9\textwidth]{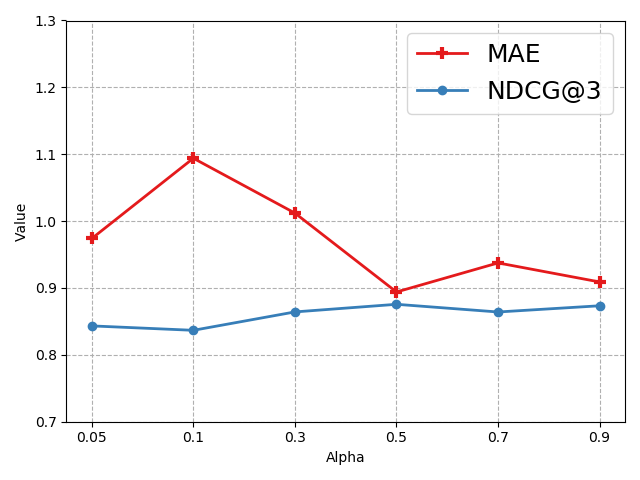}
\centering{(e)}
\end{minipage}
\begin{minipage}[t]{4cm}
\includegraphics[width=0.9\textwidth]{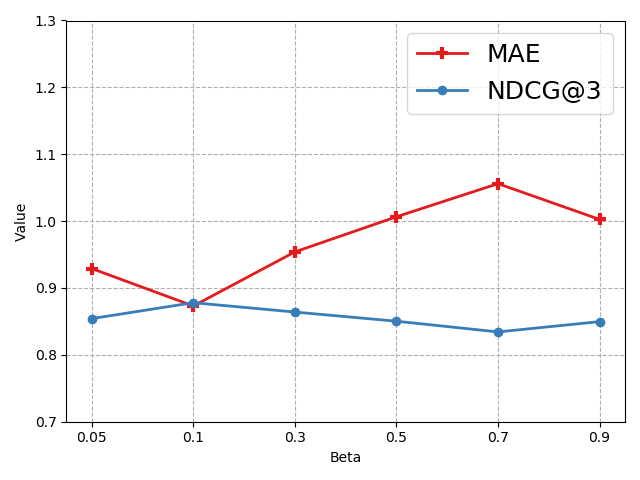}
\centering{(f)}
\end{minipage}
\begin{minipage}[t]{4cm}
\includegraphics[width=0.9\textwidth]{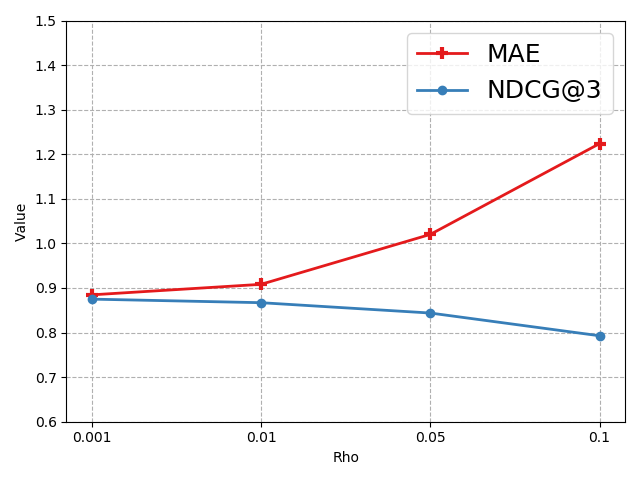}
\centering{(g)}
\end{minipage}
\begin{minipage}[t]{4.1cm}
\includegraphics[width=0.9\textwidth]{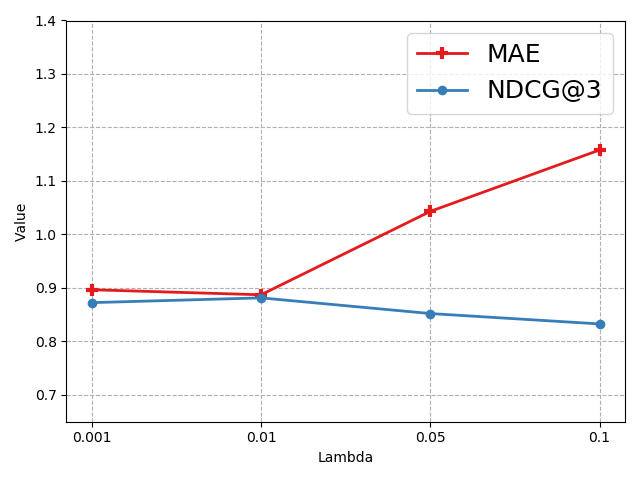}
\centering{(h)}
\end{minipage}
\caption{Parameter sensitivity in different settings. }
\label{fig:parameter_study}
\end{figure*}

\subsection{Parameter Studies}

Here we show the parameter studies on several key parameters of our method: the model set-up parameters when initializing the training and testing dataset; the model structure parameters for constructing the recommender models; and the hyper-parameters during the learning process. 
We present the results under two evaluation metrics, i.e. Mean Absolute Error ($MAE$) and normalized discounted cumulative gain ($NDCG$):
\begin{equation}
\label{eq:mae_def}
    MAE = \frac{1}{|U^{test}|}\Sigma_{u\in U^{test}}\frac{1}{|\mathcal{I}_{u}^{Q}|}\Sigma_{i\in \mathcal{I}^{Q}_{u}}|y_{u,i} - \hat{y_{u,i}}|
\end{equation}
\begin{equation}
\label{eq:ndcg_def1}
    NDCG@N = \frac{1}{|U^{test}|}\Sigma_{u\in U^{test}}\frac{DCG@N}{IDCG@N}
\end{equation}
\begin{equation}
\label{eq:ndcg_def2}
    DCG@N = \Sigma_{n=1}^{N}\frac{2^{y_{u,n}}-1}{\log(n+1)}
\end{equation}
where $N$ is the number of instances in the query set for each user; $DCG@N$ calculates the $top-N$ actual rating values sorted by predicted rating values; and $IDCG@N$ calculates the $top-N$ sorted actual rating values, i.e. the most possible values, for $DCG@N$.
$MAE$ evaluates the accuracy of rating prediction, where a lower value indicates better model performance. $NDCG@N$ considers the preference ordering performance on observed query sets, where a higher value indicates a better performance. The experiments are conducted on MovieLens dataset.

\begin{table*}[ht]
\caption{Comparison Results}
\centering
\begin{tabular}{|c|c|c|c|c|c|c|c|c|c|}
\hline
\multirow{2}{*}{Method} & \multirow{2}{*}{Metrics} & \multicolumn{4}{c|}{MovieLens-1M} & \multicolumn{4}{c|}{Book-crossing} \\ \cline{3-10}
& & W-W & W-C & C-W & C-C & W-W & W-C & C-W & C-C \\ \hline \hline
\multirow{2}{*}{MeLU}             & MAE & 0.7661 & 0.9361 & 0.7884 & 0.9299 & 0.7799 & 2.1047 & 1.8701 & 2.1475 \\ \cline{2-10}
                             & $NDCG@3$ & 0.8904 & 0.7990 & 0.8810 & 0.8011 & 0.9572 & 0.8441 & 0.8527 & 0.8410 \\ \hline
\multirow{2}{*}{MetaCS-DNN}       & MAE & 0.9047 & 1.1694 & 1.0625 & 1.2012 & 1.6206 & 2.1457 & 2.2648 & 2.3088 \\ \cline{2-10}
                             & $NDCG@3$ & 0.9090 & 0.7715 & 0.8559 & 0.7680 & 0.8860 & 0.8365 & 0.8202 & 0.8160 \\ \hline
\multirow{2}{*}{$s^{2}$ Meta}     & MAE & 0.7567 & 0.8860 & 0.8443 & 0.9130 & 1.5147 & 1.9767 & 1.8738 & 2.0921 \\ \cline{2-10}
                             & $NDCG@3$ & 0.8870 & 0.8323 & 0.8401 & 0.8148 & 0.8781 & 0.8463 & 0.8490 & 0.8422 \\ \hline
\multirow{2}{*}{Item-level RUM}   & MAE & 0.8874 & 1.3424 & 1.2655 & 1.3509 & 1.4611 & 2.2857 & 2.2608 & 2.3197 \\ \cline{2-10}
                             & $NDCG@3$ & 0.9102 & 0.7692 & 0.7420 & 0.7680 & 0.8920 & 0.8265 & 0.8202 & 0.8160 \\ \hline
\multirow{2}{*}{Feature-level RUM}& MAE & 0.8739 & 1.2439 & 1.2488 & 1.2773 & 1.3945 & 2.0837 & 2.0322 & 2.1260 \\ \cline{2-10}
                             & $NDCG@3$ & 0.9120 & 0.7721 & 0.7642 & 0.7547 & 0.8909 & 0.8430 & 0.8338 & 0.8273 \\ \hline \hline
\multirow{2}{*}{MAMO}             & MAE & 0.8725 & 0.9306 & 0.8967 & 0.8894 & 1.4879 & 1.7379 & 1.6217 & 1.8188 \\ \cline{2-10}
                             & $NDCG@3$ & 0.8866 & 0.8315 & 0.8799 & 0.8709 & 0.8879 & 0.8402 & 0.8565 & 0.8384 \\
\hline
\end{tabular}
\label{tab:comparison_results}
\end{table*}

\subsubsection{Model set-up parameters.}
Here we consider the model performance under different separating ratios for training-testing or support-query sets.
The experimental results (Figure~\ref{fig:parameter_study} (a)) suggest a higher ratio of the training set improves the model performance; while the performance for training with only half of the users is still acceptable, which indicates the effectiveness of the meta-training approach. 
As for the sample size of the support set, where we control the maximum of rating records for a user as 20, we compare the results with support sample size between 5 to 15, where we use the last 5 samples for each user as the query sample set for a fair comparison. 
According to Figure~\ref{fig:parameter_study} (b), we can find that a larger sample size provides better results, while the model can also provide acceptable predictions with smaller sample size, which shows its capacity in cold scenarios.

\subsubsection{Model structure parameters.}
The settings of the parameters for deep neural nets, especially the number of layers and the dimension of layer nodes, can sometimes affect the model performance. 
Generally, a network with large dimension and a complex structure may provide accurate predictions but is data hungry and slow in convergence. In our work, we use fully connected layers to learn the user and item embeddings. According to Figure~\ref{fig:parameter_study} (c)-(d), which show the model performance under different number of embedding layers and embedding size $d_{e}$, we could see that the number of neural layers has slight impacts on the model performance, while too shallow or too deep layers may lead to bad results. A moderate setting of the embedding size can provide acceptable prediction results and require less training time. 

\subsubsection{Hyper-parameters.}
We have hyper-parameters $\alpha$ and $\beta$ (ranging from 0 to 1) for globally updating the profile memory $M_{P}$ and user embedding memory $M_{U}$ in feature-specific memory, and $\gamma$ for globally updating the task-specific memory $M_{U,I}$. The hyper-parameters control how much new information is added to the memory matrices. Figure~\ref{fig:parameter_study} (e)-(f) show the parameter sensitivity for parameter $\alpha$ and $\beta$. For parameter $\alpha$, values around 0.5 provide better performance of the models, because of the update strategy of the memory matrices. Since the profile memory is randomly initialized and is not updated by back-propagation, the profile memory requires more new information to complete the distribution of the user profiles. While for the user embedding memory, too much new information may cause chaos in existing memory, where $\beta=0.1$ provides the best performance. The parameter $\gamma$ for task-specific memory (not listed in the figures) has similar patterns with $\beta$, where it provides best results with values around 0.1.
We have $\rho$ and $\lambda$ for updating local or global parameters, and $\tau$ for providing personalized bias term. Figure~\ref{fig:parameter_study} (g)-(h) show the results for $\rho$ and $\lambda$, where small learning rates provide best results, since a large value may make the model difficult in convergence. As for the parameter $\tau$, which is not listed in the figures, we find the values around 0.1 perform best. 

\subsection{Comparison Results}
\subsubsection{Comparison Methods}
We adopt the following representative state-of-the-arts that apply the meta-learning idea into the recommender systems, where the first three methods are based on meta-optimization, and the last method, which includes item-level model and feature-level model, is based on memory networks.
\begin{itemize}
    \item \textbf{MeLU} \cite{lee2019melu}: get rating predictions by feeding the concatenation of user and item embeddings into fully connected layers. The local parameter of the fully connected layers will be locally updated for personalized recommendation. The global parameter of the recommender and the parameters of embedding layers will be globally updated for all the users.
    \item \textbf{MetaCS-DNN} \cite{bharadhwaj2019meta}: follows a similar idea of MeLU when constructing the recommender model, i.e. several fully connected layers, the difference is the local and global parameters involve all parameters from the inputs to the predictions.
    \item \textbf{$s^{2}$ Meta} \cite{du2019sequential}: uses a deep neural net as the recommender model and updated via meta-optimization approach. Specifically, it introduced update and stop controllers to update the local parameters. 
    \item \textbf{RUM}: \cite{chen2018sequential} uses neural matrix factorization as the recommender model, where the score for an item is the product of the user and item embeddings. The user embedding is learned from a user's intrinsic embedding and a memory embedding. The memory embedding is learned by either item-level or feature-level memories, i.e. \textbf{item-level RUM} and \textbf{feature-level RUM}. The item-level RUM stores and extracts the memory according to different items, while the feature-level RUM considers the similarity of the different features. Here, the memory matrix is initialized by each user's history records and is locally updated for personalized recommendation.
\end{itemize}
The parameter settings for our method and the comparison methods can be found in appendix~\ref{sec:appendix_comparison}.

\subsubsection{Comparison Results}
The comparison results are listed in Table~\ref{tab:comparison_results}. Generally, we can see that the meta-optimization based methods show outstanding performance in cold scenarios, while RUM outperforms other methods in warm situations. This can be attributed to the stronger capability of meta-optimization approach in capturing the general preferences. Since the memory matrices designed in RUM store only the personal history information, it needs more history records to establish an efficient memory matrix. Comparing MetaCS-DNN and MeLU, which utilize different strategies for updating embedding parameters, the results show that locally updating the embedding parameters performs better than globally updating the embedding parameters. We can see that our method shows stable and well performance in different scenarios, which shows the effectiveness of the meta-augmented meta-optimization strategy: learning a personalized initialization of local parameters from the memories. 

\subsection{Discussion}
MAMO has two groups of memories. The first group includes user profile memory $M_{P}$ and user embedding memory $M_{U}$ that are utilized for providing personalized bias term when initializing the local parameters. The second group includes the task-specific memory $M_{U,I}$.
In this section, we will discuss the impact of these two groups of memories. Also, we predefined $K$ user types when constructing the aforementioned memories. We will discuss how to set an appropriate $K$ and present an example of the standard users for $K$ user preference patterns.

\begin{figure}[t]
\centering
\begin{minipage}[t]{3.6cm}
\includegraphics[width=\textwidth]{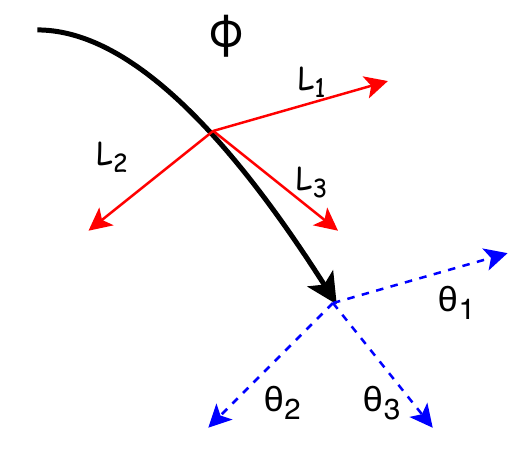}
\centering{(a)}
\end{minipage}
\begin{minipage}[t]{3.8cm}
\includegraphics[width=\textwidth]{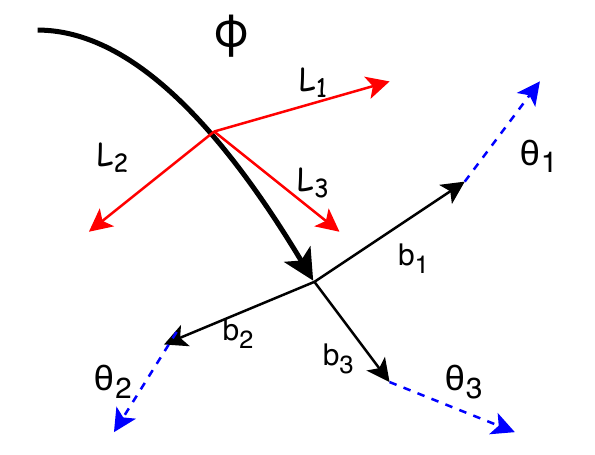}
\centering{(b)}
\end{minipage}
\caption{The major difference between MAML and our method. Red lines stand for the global/meta-optimization, the blue lines stand for the local learning/adaption.}
\label{fig:ablation_study1}
\end{figure}
\subsubsection{Impact of Personalized Bias Term}
The basic idea of meta-optimization process is learning a sharing global parameter $\phi$ when initializing the personalized local parameter $\theta$ (see Figure~\ref{fig:ablation_study1} (a)). For addressing the potential local optima issues by the unique global sharing parameter for most current meta-optimization based works, we introduce two memory matrices to provide a personalized bias term $b_{u}$ when initializing the local parameters $\theta_{u}$ when learning the user embedding. Specifically, the profile memory $M_{P}$ stores $K$ types of user profiles, and $M_{U}$ stores corresponding bias terms. Then, the local parameters will be locally updated for personalized recommendation (see Figure~\ref{fig:ablation_study1} (b)). The comparison methods, MeLU and MetaCS-DNN, both utilize MAML for meta-optimization. According to the comparison results, we can see that our strategy outperforms these two methods in cold scenarios, especially for cold users. The reason may lie in that our optimization approach takes the user profile into consideration. The profile memory will automatically cluster users with similar profiles to provide fast bias terms.

\begin{figure}[t]
\centering
\begin{minipage}[t]{4cm}
\includegraphics[width=0.9\textwidth]{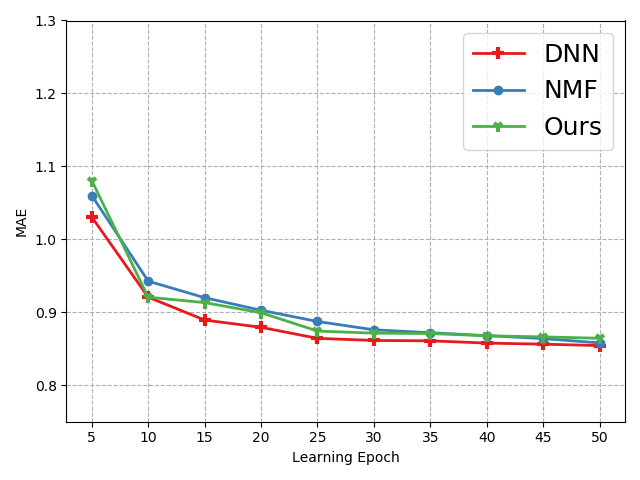}
\centering{(a)Training: |S|=10}
\end{minipage}
\begin{minipage}[t]{4cm}
\includegraphics[width=0.9\textwidth]{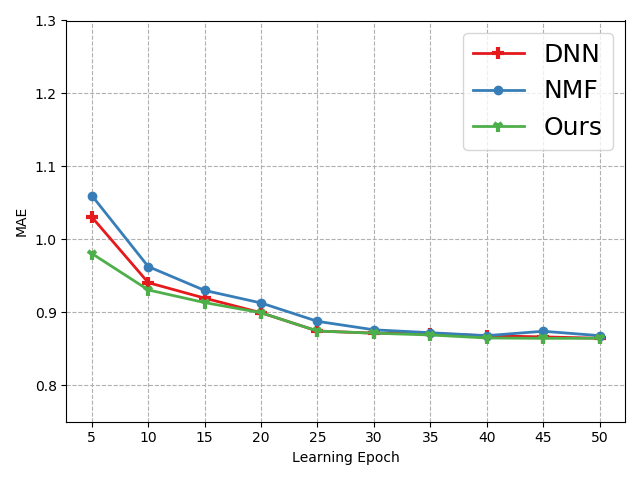}
\centering{(b)Testing: |S|=10}
\end{minipage}
\begin{minipage}[t]{4cm}
\includegraphics[width=0.9\textwidth]{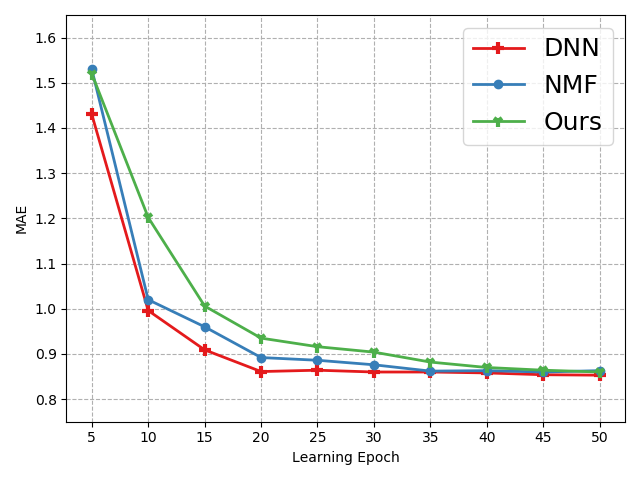}
\centering{(c)Training: |S|=5}
\end{minipage}
\begin{minipage}[t]{4cm}
\includegraphics[width=0.9\textwidth]{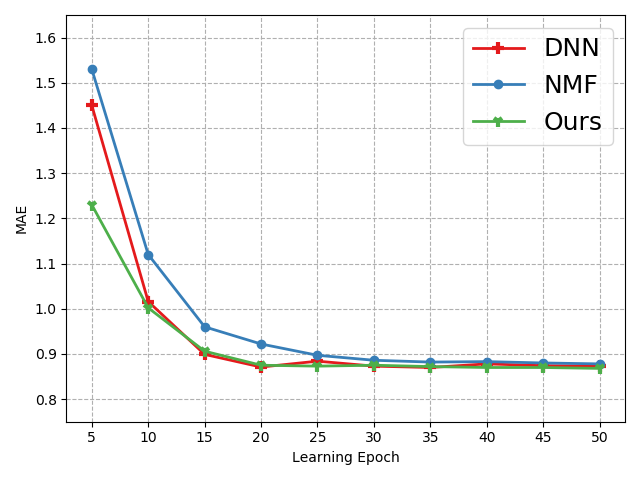}
\centering{(d)Testing: |S|=5}
\end{minipage}
\caption{The performance during training process and testing process with different settings of support sample size on different recommender models.}
\label{fig:ablation_study2}
\end{figure}

\subsubsection{Impact of Preference Memory}
We also introduce the task-specific memory $M_{U,I}$ that contains $K$ types of preference matrices during the recommendation process. To see whether it helps the recommendation process, we compare it with different designs of recommender models. The first is predicting the ratings by the concatenation of the user and item embedding through several fully connected layers, denoted by DNN. The second is taking the idea of neural matrix factorization~\cite{he2017neural}, which multiplies the user embedding and item embedding as the predictions, denoted by NMF. The results are shown in Figure~\ref{fig:ablation_study2}, where (a), (c) show the performance on the query dataset during the training process; (b), (d) show the performance on the query dataset during the testing process. We can observe that DNN and NMF converge fast in the training process but perform unwell on testing dataset with small support sample size, while our method needs more steps to store the preference information but can learn fast during the testing process.

\subsubsection{Discussion of Preference Type}
\begin{figure}[t]
\centering
\begin{minipage}[t]{4cm}
\includegraphics[width=0.9\textwidth]{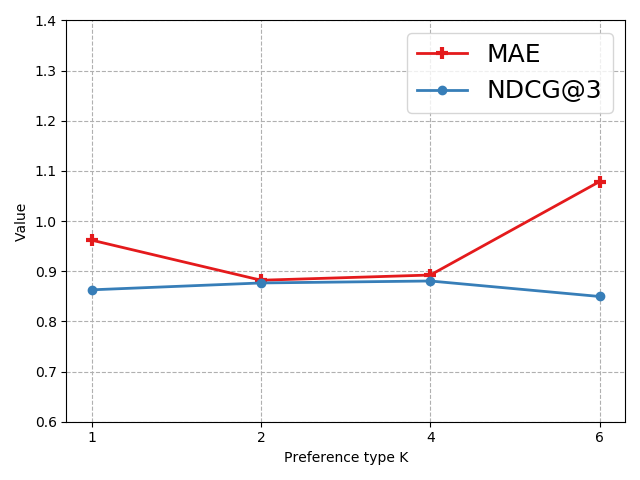}
\centering{(a)}
\end{minipage}
\begin{minipage}[t]{4cm}
\includegraphics[width=0.9\textwidth]{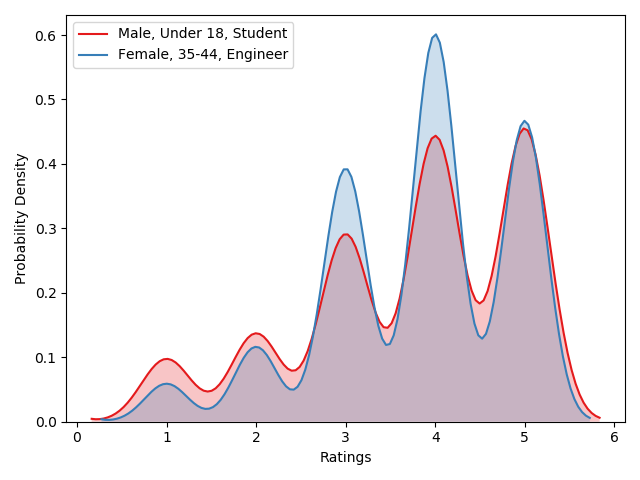}
\centering{(b)}
\end{minipage}
\caption{(a)Performance with different $K$. (b) Example for representative users when $K=2$.}
\label{fig:ablation_study3}
\end{figure}

In this work we predefined $K$ user types when constructing the proposed memories, which can be regarded as the number of clusters of users based on the user profiles. 
Figure~\ref{fig:ablation_study3} (a) shows the model performance under different $K$. We can see that the $K$ values around 2-4 achieve the best performance. A larger $K$ has potential in providing more accurate guidance in recommendation, but it will take more computation cost (each type in $M_{U}$ stores parameters with the same shape as all the network parameters in user embedding networks). In Figure~\ref{fig:ablation_study3} (b), we present the case when $K=2$. We selected two representative user profiles which are most similar to $M_{P}$, the distributions are from the top 100 users that belong to these two types. 

\subsubsection{Limitations and Future work}
In cold-start scenarios, auxiliary information is valuable for providing potential recommendations. Our work targets at the situations that users or items have limited interaction histories and utilizes the idea of few-shot learning to provide recommendations. We assume we have enough essential profile details of the users or items. During the data preprocessing process, we filter the data that are without relevant profiles. However, in the real-world cases, the auxiliary information is not always attainable, where our model becomes inefficient under such circumstance.
In future works, we plan to leverage the knowledge from multi-modal studies~\cite{baltruvsaitis2018multimodal} to address this issue. For example, Wu et al. \cite{wu2018multimodal} design a multimodal variational autoencoder that learns a joint feature distribution under any missing modalities.

\section{Related Work}
\label{sec:sec_4}

\subsection{Cold-start Problem in Recommender Systems}
It is still a tricky and significant problem for recommenders to accurately recommend items to users. Some solutions in solving this problem have been provided in current publications \cite{mishra2017tools}, involving second-domain knowledge transfer \cite{li2018cross}, auxiliary and contextual information \cite{roy2016latent,yao2018collaborative}, active learning \cite{zhu2019addressing} and deep learning \cite{ebesu2017neural}.

Commonly, it is reported that the more information the better recommendation results are. \cite{mirbakhsh2015improving} adds the matrix factorization with clusters to recommendations of cross-domains. Similarly, \cite{li2018cross} presents an innovative model of cross-domain recommendation according to the partial least squares regression (PLSR) analysis. Both PLSR-CrossRec and PLSR-Latent can be utilized in rating of source-domain for better predicting ratings of cold-start users. Apart from leveraging auxiliary information from another domain, some researchers have regarded representative item content as auxiliary information to learn representative latent factor when coped with the cold-start problem. According to implied feedback, \cite{roy2016latent} presents an approach named visual-CLiMF to learn representative latent factors for cold start videos, where emotional aspects of items are incorporated into the latent factor representations of video contents. In order to better use content characteristics in further latent characteristics, \cite{chou2016addressing} proposed a next-song recommender system for cold-start recommendation by mapping both learning and updating within spaces of the audio characteristic and the item latent. In addition, active learning has found its way in tackling both the user cold-start problem in both users \cite{elahi2016survey} and the items \cite{anava2015budget}. \cite{zhu2019addressing} combines the active learning approach with item’s attribute information to deal with the cold-start problem in items. More recent works take advantage of deep learning \cite{wei2016collaborative,ebesu2017neural,yuan2016solving}. \cite{ma2020deep} proposes a multiplex interaction-oriented service recommendation approach (MISR), which merged different interactions into a deep neural network to search latent information so that new users can be better dealt with. \cite{wei2016collaborative} proposes a hybrid recommendation model in researching the content characteristics of items by means of a deep learning neural network and then exploited them to the timeSVD++ collaborative filtering model.

\subsection{Meta-learning for Recommender Systems}
Meta-learning, namely learning to learn, is learning across various tasks with fewer training samples in each task, which can be easily adapted to new tasks \cite{vanschoren2018meta}. 
In recent years, meta-learning has been attracting much attention to alleviate cold-start problems~\cite{chen2018federated,lee2019melu,zhao2019learning}, and most of them adopted optimization-based meta-learning approach and chose model-agnostic meta-learning (MAML)~\cite{finn2017model} for model training.
Generally, those works define a recommendation model $\mathcal{R}_{\theta}$ with parameter $\theta$, and a meta-learner $\mathcal{M}(\phi)$. Modeling each user's preference is regarded as a learning task, and the users are divided into training and testing users.
The initialization of parameter $\theta$ is defined by some function $\theta \leftarrow \mathcal{M}({\phi})$, e.g. in a simplest case, $\theta \leftarrow \phi$. During the training process, the parameter of the recommendation model, i.e. $\theta$, will be locally updated by minimizing the recommendation loss $\mathcal{L}(\mathcal{R}_{\theta}(D^{support}))$ on the support set $D^{support}$ (the history ratings by a user) for a single user; and then used to retrieve the recommendation loss $\mathcal{L}(\mathcal{R}_{\theta}(D^{query}))$ on the query set $D^{query}$. The parameter of the meta-learner will be globally updated by minimizing the sum of recommendation loss $\Sigma_{u\in U^{train}}(\mathcal{L}(\mathcal{R}_{\theta \leftarrow \phi}(D^{query})))$ for all users in the training dataset. Then, the meta-learner will guide the learning process for new users by providing the initialization of parameter $\theta$, which can expedite the learning process of the recommendation models for testing users.

The major differences among the above works lie in the design of the recommender model $\mathcal{R}_{\theta}$ and the meta-optimization approach $\mathcal{M}({\phi})$. For example, \cite{lee2019melu} targets at a rating prediction problem where they feed the concatenation of user and item embedding into fully-connected layers to get predictions. They locally update the parameters of neural layers to get personalized recommendation and globally update the parameters of embedding learning and a global parameter of neural layers. \cite{du2019sequential} constructs a deep neural nets as the recommender model, and they design an input gate and a forget gate during updating the local parameters.
One limitation of the current meta-optimization approach for recommendation is that most works learn a shared global initialization parameter, which means the initialization parameter is same for all users. This may lead the recommender model to the local optima problem and to be slow in convergence. 
To address this issue, our work introduces two memory matrices to provide personalized bias terms when initializing the parameters.

\subsection{Memory-augmented Neural Networks}
With the ability to express, store, and manipulate the records explicitly, dynamically, and effectively, external memory networks \cite{sukhbaatar2015end} have gained popularity in recent years in fields such as question-answering systems \cite{xiong2016dynamic} and knowledge tracking~\cite{ghazvininejad2018knowledge}.
One branch of the memory neural networks is based on Neural Turing Machine (NTM) \cite{graves2014neural}: the memory is normally stored in a matrix with a read head to extract the information from the memory and a write head to update the memory at each time step. Such features make the NTM a good module for meta-learning or few-shot learning.
Following this idea, \cite{santoro2016meta} proposes Memory-augmented Neural Networks (MANN). Similar to NTM, they design a read controller to learn a key vector $v_{t}$ from the inputs at time $t$, and $v_{t}$ is used to learn the rating weights $w^{r}_{t}$ for the rows in the memory matrix $M_{t}$; the retrieved vector is the sum of rows in memory matrix with the above attention values, i.e. $\Sigma_{k}w^{r}_{t}(k)M_{t}(k)$.
Then, for updating the memory, the authors propose usage weights $w^{u}_{t}$ to evaluate the usage of the rows in the memory matrix. The writing weights $w^{w}_{t}$ are learned from the previous reading weights and the usage weights. Last, the memory is updated by $M_{t}(k) \leftarrow M_{t-1}(k) + w^{w}_{t}(k)v_{t}$.

The work by \cite{chen2018sequential} is one of the first works that integrate MANN into recommendation tasks. They consider the matrix factorization model as a neural network, where the score $y_{u,i}$ for an item $i$ from user $u$ is predicted from the user and item embeddings, i.e. $y_{u,i}=e_{u}^{\top}e_{i}$. They propose memory enhanced the user embedding, where the user embedding is learned from a user's intrinsic embedding $e^{*}_{u}$ and a memory embedding $e^{M}_{u}=read(M_{u}, e_{i})$, i.e. $e_{u} \leftarrow f(e^{*}_{u}, e^{M}_{u})$. The memory embedding $e^{M}_{u}$ reads from the user personalized memory matrix $M_{u}$ to retrieve the related information for item $i$ according to the item embedding $e_{i}$. Another work by \cite{li2019gated} adds a gate mechanism to control how much past memory and current memory are kept.
Different from these works where the memory matrices are designed for each user and cannot share among different users, we propose two global sharing memories: feature-specific memory and task-specific memory. 
The feature-specific memory includes user profile memory and user embedding memory.
Given a user, this memory will extract the parameter memory according to the similarity between the user profile and the user profile memory. 
The task-specific memory stores the preference information, which serves as fast weights for the recommender model, to alleviate the need for storing copies of neural activity patterns.

\section{Conclusion}
\label{sec:sec_5}
A common challenge for most current recommender systems is the cold-start problem. Recently, inspired by the progress of few-shot learning, some works leverage the meta-optimization idea into the recommendation. 
The core idea is learning a global sharing initialization parameter of the recommender model for all users and then updating the local parameters to learn personalized recommender model for single users. 
A common limitation for most current works is that the global sharing initialization parameter is unique for all users, which may lead the recommendation model into local optima for users with distinct preference patterns.
For addressing this issue, in this work, we design two memory matrices to provide personalized initialization for recommender models: feature-specific memory that provides a personalized bias term when initializing the recommender model, and a task-specific memory to guide the recommendation process.
We evaluate the proposed methods on two public available datasets and provide details of the implementation. The experimental results show the effectiveness of our method. 

\bibliographystyle{ACM-Reference-Format}
\bibliography{acmart}


\begin{thebibliography}{35}


\ifx \showCODEN    \undefined \def \showCODEN     #1{\unskip}     \fi
\ifx \showDOI      \undefined \def \showDOI       #1{#1}\fi
\ifx \showISBNx    \undefined \def \showISBNx     #1{\unskip}     \fi
\ifx \showISBNxiii \undefined \def \showISBNxiii  #1{\unskip}     \fi
\ifx \showISSN     \undefined \def \showISSN      #1{\unskip}     \fi
\ifx \showLCCN     \undefined \def \showLCCN      #1{\unskip}     \fi
\ifx \shownote     \undefined \def \shownote      #1{#1}          \fi
\ifx \showarticletitle \undefined \def \showarticletitle #1{#1}   \fi
\ifx \showURL      \undefined \def \showURL       {\relax}        \fi
\providecommand\bibfield[2]{#2}
\providecommand\bibinfo[2]{#2}
\providecommand\natexlab[1]{#1}
\providecommand\showeprint[2][]{arXiv:#2}

\bibitem[\protect\citeauthoryear{Anava, Golan, Golbandi, Karnin, Lempel,
  Rokhlenko, and Somekh}{Anava et~al\mbox{.}}{2015}]%
        {anava2015budget}
\bibfield{author}{\bibinfo{person}{Oren Anava}, \bibinfo{person}{Shahar Golan},
  \bibinfo{person}{Nadav Golbandi}, \bibinfo{person}{Zohar Karnin},
  \bibinfo{person}{Ronny Lempel}, \bibinfo{person}{Oleg Rokhlenko}, {and}
  \bibinfo{person}{Oren Somekh}.} \bibinfo{year}{2015}\natexlab{}.
\newblock \showarticletitle{Budget-constrained item cold-start handling in
  collaborative filtering recommenders via optimal design}. In
  \bibinfo{booktitle}{\emph{Proceedings of the 24th International Conference on
  World Wide Web}}. \bibinfo{pages}{45--54}.
\newblock


\bibitem[\protect\citeauthoryear{Baltru{\v{s}}aitis, Ahuja, and
  Morency}{Baltru{\v{s}}aitis et~al\mbox{.}}{2018}]%
        {baltruvsaitis2018multimodal}
\bibfield{author}{\bibinfo{person}{Tadas Baltru{\v{s}}aitis},
  \bibinfo{person}{Chaitanya Ahuja}, {and} \bibinfo{person}{Louis-Philippe
  Morency}.} \bibinfo{year}{2018}\natexlab{}.
\newblock \showarticletitle{Multimodal machine learning: A survey and
  taxonomy}.
\newblock \bibinfo{journal}{\emph{IEEE transactions on pattern analysis and
  machine intelligence}} \bibinfo{volume}{41}, \bibinfo{number}{2}
  (\bibinfo{year}{2018}), \bibinfo{pages}{423--443}.
\newblock


\bibitem[\protect\citeauthoryear{Bharadhwaj}{Bharadhwaj}{2019}]%
        {bharadhwaj2019meta}
\bibfield{author}{\bibinfo{person}{Homanga Bharadhwaj}.}
  \bibinfo{year}{2019}\natexlab{}.
\newblock \showarticletitle{Meta-Learning for User Cold-Start Recommendation}.
  In \bibinfo{booktitle}{\emph{2019 International Joint Conference on Neural
  Networks (IJCNN)}}. IEEE, \bibinfo{pages}{1--8}.
\newblock


\bibitem[\protect\citeauthoryear{Bokde, Girase, and Mukhopadhyay}{Bokde
  et~al\mbox{.}}{2015}]%
        {bokde2015matrix}
\bibfield{author}{\bibinfo{person}{Dheeraj Bokde}, \bibinfo{person}{Sheetal
  Girase}, {and} \bibinfo{person}{Debajyoti Mukhopadhyay}.}
  \bibinfo{year}{2015}\natexlab{}.
\newblock \showarticletitle{Matrix factorization model in collaborative
  filtering algorithms: A survey}.
\newblock \bibinfo{journal}{\emph{Procedia Computer Science}}
  \bibinfo{volume}{49} (\bibinfo{year}{2015}), \bibinfo{pages}{136--146}.
\newblock


\bibitem[\protect\citeauthoryear{Chen, Dong, Li, and He}{Chen
  et~al\mbox{.}}{2018a}]%
        {chen2018federated}
\bibfield{author}{\bibinfo{person}{Fei Chen}, \bibinfo{person}{Zhenhua Dong},
  \bibinfo{person}{Zhenguo Li}, {and} \bibinfo{person}{Xiuqiang He}.}
  \bibinfo{year}{2018}\natexlab{a}.
\newblock \showarticletitle{Federated meta-learning for recommendation}.
\newblock \bibinfo{journal}{\emph{arXiv preprint arXiv:1802.07876}}
  (\bibinfo{year}{2018}).
\newblock


\bibitem[\protect\citeauthoryear{Chen, Xu, Zhang, Tang, Cao, Qin, and Zha}{Chen
  et~al\mbox{.}}{2018b}]%
        {chen2018sequential}
\bibfield{author}{\bibinfo{person}{Xu Chen}, \bibinfo{person}{Hongteng Xu},
  \bibinfo{person}{Yongfeng Zhang}, \bibinfo{person}{Jiaxi Tang},
  \bibinfo{person}{Yixin Cao}, \bibinfo{person}{Zheng Qin}, {and}
  \bibinfo{person}{Hongyuan Zha}.} \bibinfo{year}{2018}\natexlab{b}.
\newblock \showarticletitle{Sequential recommendation with user memory
  networks}. In \bibinfo{booktitle}{\emph{Proceedings of the eleventh ACM
  international conference on web search and data mining}}. ACM,
  \bibinfo{pages}{108--116}.
\newblock


\bibitem[\protect\citeauthoryear{Chou, Yang, Jang, and Lin}{Chou
  et~al\mbox{.}}{2016}]%
        {chou2016addressing}
\bibfield{author}{\bibinfo{person}{Szu-Yu Chou}, \bibinfo{person}{Yi-Hsuan
  Yang}, \bibinfo{person}{Jyh-Shing~Roger Jang}, {and}
  \bibinfo{person}{Yu-Ching Lin}.} \bibinfo{year}{2016}\natexlab{}.
\newblock \showarticletitle{Addressing cold start for next-song
  recommendation}. In \bibinfo{booktitle}{\emph{Proceedings of the 10th ACM
  Conference on Recommender Systems}}. \bibinfo{pages}{115--118}.
\newblock


\bibitem[\protect\citeauthoryear{Du, Wang, Yang, Zhou, and Tang}{Du
  et~al\mbox{.}}{2019}]%
        {du2019sequential}
\bibfield{author}{\bibinfo{person}{Zhengxiao Du}, \bibinfo{person}{Xiaowei
  Wang}, \bibinfo{person}{Hongxia Yang}, \bibinfo{person}{Jingren Zhou}, {and}
  \bibinfo{person}{Jie Tang}.} \bibinfo{year}{2019}\natexlab{}.
\newblock \showarticletitle{Sequential Scenario-Specific Meta Learner for
  Online Recommendation}.
\newblock \bibinfo{journal}{\emph{arXiv preprint arXiv:1906.00391}}
  (\bibinfo{year}{2019}).
\newblock


\bibitem[\protect\citeauthoryear{Ebesu and Fang}{Ebesu and Fang}{2017}]%
        {ebesu2017neural}
\bibfield{author}{\bibinfo{person}{Travis Ebesu} {and} \bibinfo{person}{Yi
  Fang}.} \bibinfo{year}{2017}\natexlab{}.
\newblock \showarticletitle{Neural Semantic Personalized Ranking for item
  cold-start recommendation}.
\newblock \bibinfo{journal}{\emph{Information Retrieval Journal}}
  \bibinfo{volume}{20}, \bibinfo{number}{2} (\bibinfo{year}{2017}),
  \bibinfo{pages}{109--131}.
\newblock


\bibitem[\protect\citeauthoryear{Elahi, Ricci, and Rubens}{Elahi
  et~al\mbox{.}}{2016}]%
        {elahi2016survey}
\bibfield{author}{\bibinfo{person}{Mehdi Elahi}, \bibinfo{person}{Francesco
  Ricci}, {and} \bibinfo{person}{Neil Rubens}.}
  \bibinfo{year}{2016}\natexlab{}.
\newblock \showarticletitle{A survey of active learning in collaborative
  filtering recommender systems}.
\newblock \bibinfo{journal}{\emph{Computer Science Review}}
  \bibinfo{volume}{20} (\bibinfo{year}{2016}), \bibinfo{pages}{29--50}.
\newblock


\bibitem[\protect\citeauthoryear{Finn, Abbeel, and Levine}{Finn
  et~al\mbox{.}}{2017}]%
        {finn2017model}
\bibfield{author}{\bibinfo{person}{Chelsea Finn}, \bibinfo{person}{Pieter
  Abbeel}, {and} \bibinfo{person}{Sergey Levine}.}
  \bibinfo{year}{2017}\natexlab{}.
\newblock \showarticletitle{Model-agnostic meta-learning for fast adaptation of
  deep networks}. In \bibinfo{booktitle}{\emph{Proceedings of the 34th
  International Conference on Machine Learning-Volume 70}}. JMLR. org,
  \bibinfo{pages}{1126--1135}.
\newblock


\bibitem[\protect\citeauthoryear{Ghazvininejad, Brockett, Chang, Dolan, Gao,
  Yih, and Galley}{Ghazvininejad et~al\mbox{.}}{2018}]%
        {ghazvininejad2018knowledge}
\bibfield{author}{\bibinfo{person}{Marjan Ghazvininejad},
  \bibinfo{person}{Chris Brockett}, \bibinfo{person}{Ming-Wei Chang},
  \bibinfo{person}{Bill Dolan}, \bibinfo{person}{Jianfeng Gao},
  \bibinfo{person}{Wen-tau Yih}, {and} \bibinfo{person}{Michel Galley}.}
  \bibinfo{year}{2018}\natexlab{}.
\newblock \showarticletitle{A knowledge-grounded neural conversation model}. In
  \bibinfo{booktitle}{\emph{Thirty-Second AAAI Conference on Artificial
  Intelligence}}.
\newblock


\bibitem[\protect\citeauthoryear{Graves, Wayne, and Danihelka}{Graves
  et~al\mbox{.}}{2014}]%
        {graves2014neural}
\bibfield{author}{\bibinfo{person}{Alex Graves}, \bibinfo{person}{Greg Wayne},
  {and} \bibinfo{person}{Ivo Danihelka}.} \bibinfo{year}{2014}\natexlab{}.
\newblock \showarticletitle{Neural turing machines}.
\newblock \bibinfo{journal}{\emph{arXiv preprint arXiv:1410.5401}}
  (\bibinfo{year}{2014}).
\newblock


\bibitem[\protect\citeauthoryear{He, Liao, Zhang, Nie, Hu, and Chua}{He
  et~al\mbox{.}}{2017}]%
        {he2017neural}
\bibfield{author}{\bibinfo{person}{Xiangnan He}, \bibinfo{person}{Lizi Liao},
  \bibinfo{person}{Hanwang Zhang}, \bibinfo{person}{Liqiang Nie},
  \bibinfo{person}{Xia Hu}, {and} \bibinfo{person}{Tat-Seng Chua}.}
  \bibinfo{year}{2017}\natexlab{}.
\newblock \showarticletitle{Neural collaborative filtering}. In
  \bibinfo{booktitle}{\emph{Proceedings of the 26th international conference on
  world wide web}}. \bibinfo{pages}{173--182}.
\newblock


\bibitem[\protect\citeauthoryear{Lee, Im, Jang, Cho, and Chung}{Lee
  et~al\mbox{.}}{2019}]%
        {lee2019melu}
\bibfield{author}{\bibinfo{person}{Hoyeop Lee}, \bibinfo{person}{Jinbae Im},
  \bibinfo{person}{Seongwon Jang}, \bibinfo{person}{Hyunsouk Cho}, {and}
  \bibinfo{person}{Sehee Chung}.} \bibinfo{year}{2019}\natexlab{}.
\newblock \showarticletitle{MeLU: Meta-Learned User Preference Estimator for
  Cold-Start Recommendation}. In \bibinfo{booktitle}{\emph{Proceedings of the
  25th ACM SIGKDD International Conference on Knowledge Discovery \& Data
  Mining}}. ACM, \bibinfo{pages}{1073--1082}.
\newblock


\bibitem[\protect\citeauthoryear{Li, Hsu, and Shan}{Li et~al\mbox{.}}{2018}]%
        {li2018cross}
\bibfield{author}{\bibinfo{person}{Cheng-Te Li}, \bibinfo{person}{Chia-Tai
  Hsu}, {and} \bibinfo{person}{Man-Kwan Shan}.}
  \bibinfo{year}{2018}\natexlab{}.
\newblock \showarticletitle{A Cross-Domain Recommendation Mechanism for
  Cold-Start Users Based on Partial Least Squares Regression}.
\newblock \bibinfo{journal}{\emph{ACM Transactions on Intelligent Systems and
  Technology (TIST)}} \bibinfo{volume}{9}, \bibinfo{number}{6}
  (\bibinfo{year}{2018}), \bibinfo{pages}{1--26}.
\newblock


\bibitem[\protect\citeauthoryear{Li, Song, Li, and Liu}{Li
  et~al\mbox{.}}{2019}]%
        {li2019gated}
\bibfield{author}{\bibinfo{person}{Yunxiao Li}, \bibinfo{person}{Jiaxing Song},
  \bibinfo{person}{Xiao Li}, {and} \bibinfo{person}{Weidong Liu}.}
  \bibinfo{year}{2019}\natexlab{}.
\newblock \showarticletitle{Gated Sequential Recommendation with Dynamic Memory
  Network}. In \bibinfo{booktitle}{\emph{2019 International Joint Conference on
  Neural Networks (IJCNN)}}. IEEE, \bibinfo{pages}{1--8}.
\newblock


\bibitem[\protect\citeauthoryear{Ma, Geng, and Wang}{Ma et~al\mbox{.}}{2020}]%
        {ma2020deep}
\bibfield{author}{\bibinfo{person}{Yutao Ma}, \bibinfo{person}{Xiao Geng},
  {and} \bibinfo{person}{Jian Wang}.} \bibinfo{year}{2020}\natexlab{}.
\newblock \showarticletitle{A Deep Neural Network With Multiplex Interactions
  for Cold-Start Service Recommendation}.
\newblock \bibinfo{journal}{\emph{IEEE Transactions on Engineering Management}}
  (\bibinfo{year}{2020}).
\newblock


\bibitem[\protect\citeauthoryear{Mirbakhsh and Ling}{Mirbakhsh and
  Ling}{2015}]%
        {mirbakhsh2015improving}
\bibfield{author}{\bibinfo{person}{Nima Mirbakhsh} {and}
  \bibinfo{person}{Charles~X Ling}.} \bibinfo{year}{2015}\natexlab{}.
\newblock \showarticletitle{Improving top-n recommendation for cold-start users
  via cross-domain information}.
\newblock \bibinfo{journal}{\emph{ACM Transactions on Knowledge Discovery from
  Data (TKDD)}} \bibinfo{volume}{9}, \bibinfo{number}{4}
  (\bibinfo{year}{2015}), \bibinfo{pages}{1--19}.
\newblock


\bibitem[\protect\citeauthoryear{Mishra, Mishra, and Chaturvedi}{Mishra
  et~al\mbox{.}}{2017}]%
        {mishra2017tools}
\bibfield{author}{\bibinfo{person}{Nitin Mishra}, \bibinfo{person}{Vimal
  Mishra}, {and} \bibinfo{person}{Saumya Chaturvedi}.}
  \bibinfo{year}{2017}\natexlab{}.
\newblock \showarticletitle{Tools and techniques for solving cold start
  recommendation}. In \bibinfo{booktitle}{\emph{Proceedings of the 1st
  International Conference on Internet of Things and Machine Learning}}.
  \bibinfo{pages}{1--6}.
\newblock


\bibitem[\protect\citeauthoryear{Roy and Guntuku}{Roy and Guntuku}{2016}]%
        {roy2016latent}
\bibfield{author}{\bibinfo{person}{Sujoy Roy} {and}
  \bibinfo{person}{Sharath~Chandra Guntuku}.} \bibinfo{year}{2016}\natexlab{}.
\newblock \showarticletitle{Latent factor representations for cold-start video
  recommendation}. In \bibinfo{booktitle}{\emph{Proceedings of the 10th ACM
  conference on recommender systems}}. \bibinfo{pages}{99--106}.
\newblock


\bibitem[\protect\citeauthoryear{Santoro, Bartunov, Botvinick, Wierstra, and
  Lillicrap}{Santoro et~al\mbox{.}}{2016}]%
        {santoro2016meta}
\bibfield{author}{\bibinfo{person}{Adam Santoro}, \bibinfo{person}{Sergey
  Bartunov}, \bibinfo{person}{Matthew Botvinick}, \bibinfo{person}{Daan
  Wierstra}, {and} \bibinfo{person}{Timothy Lillicrap}.}
  \bibinfo{year}{2016}\natexlab{}.
\newblock \showarticletitle{Meta-learning with memory-augmented neural
  networks}. In \bibinfo{booktitle}{\emph{International conference on machine
  learning}}. \bibinfo{pages}{1842--1850}.
\newblock


\bibitem[\protect\citeauthoryear{Sukhbaatar, Weston, Fergus,
  et~al\mbox{.}}{Sukhbaatar et~al\mbox{.}}{2015}]%
        {sukhbaatar2015end}
\bibfield{author}{\bibinfo{person}{Sainbayar Sukhbaatar},
  \bibinfo{person}{Jason Weston}, \bibinfo{person}{Rob Fergus},
  {et~al\mbox{.}}} \bibinfo{year}{2015}\natexlab{}.
\newblock \showarticletitle{End-to-end memory networks}. In
  \bibinfo{booktitle}{\emph{Advances in neural information processing
  systems}}. \bibinfo{pages}{2440--2448}.
\newblock


\bibitem[\protect\citeauthoryear{Vanschoren}{Vanschoren}{2018}]%
        {vanschoren2018meta}
\bibfield{author}{\bibinfo{person}{Joaquin Vanschoren}.}
  \bibinfo{year}{2018}\natexlab{}.
\newblock \showarticletitle{Meta-learning: A survey}.
\newblock \bibinfo{journal}{\emph{arXiv preprint arXiv:1810.03548}}
  (\bibinfo{year}{2018}).
\newblock


\bibitem[\protect\citeauthoryear{Wang, Peng, Wang, Philip, Fu, Xu, and
  Hong}{Wang et~al\mbox{.}}{2019}]%
        {wang2019cdlfm}
\bibfield{author}{\bibinfo{person}{Xinghua Wang}, \bibinfo{person}{Zhaohui
  Peng}, \bibinfo{person}{Senzhang Wang}, \bibinfo{person}{S~Yu Philip},
  \bibinfo{person}{Wenjing Fu}, \bibinfo{person}{Xiaokang Xu}, {and}
  \bibinfo{person}{Xiaoguang Hong}.} \bibinfo{year}{2019}\natexlab{}.
\newblock \showarticletitle{CDLFM: cross-domain recommendation for cold-start
  users via latent feature mapping}.
\newblock \bibinfo{journal}{\emph{Knowledge and Information Systems}}
  (\bibinfo{year}{2019}), \bibinfo{pages}{1--28}.
\newblock


\bibitem[\protect\citeauthoryear{Wang and Yao}{Wang and Yao}{2019}]%
        {wang2019few}
\bibfield{author}{\bibinfo{person}{Yaqing Wang} {and} \bibinfo{person}{Quanming
  Yao}.} \bibinfo{year}{2019}\natexlab{}.
\newblock \showarticletitle{Few-shot learning: A survey}.
\newblock \bibinfo{journal}{\emph{arXiv preprint arXiv:1904.05046}}
  (\bibinfo{year}{2019}).
\newblock


\bibitem[\protect\citeauthoryear{Wei, He, Chen, Zhou, and Tang}{Wei
  et~al\mbox{.}}{2016}]%
        {wei2016collaborative}
\bibfield{author}{\bibinfo{person}{Jian Wei}, \bibinfo{person}{Jianhua He},
  \bibinfo{person}{Kai Chen}, \bibinfo{person}{Yi Zhou}, {and}
  \bibinfo{person}{Zuoyin Tang}.} \bibinfo{year}{2016}\natexlab{}.
\newblock \showarticletitle{Collaborative filtering and deep learning based
  hybrid recommendation for cold start problem}. In
  \bibinfo{booktitle}{\emph{2016 IEEE 14th Intl Conf on Dependable, Autonomic
  and Secure Computing, 14th Intl Conf on Pervasive Intelligence and Computing,
  2nd Intl Conf on Big Data Intelligence and Computing and Cyber Science and
  Technology Congress (DASC/PiCom/DataCom/CyberSciTech)}}. IEEE,
  \bibinfo{pages}{874--877}.
\newblock


\bibitem[\protect\citeauthoryear{Wei, He, Chen, Zhou, and Tang}{Wei
  et~al\mbox{.}}{2017}]%
        {wei2017collaborative}
\bibfield{author}{\bibinfo{person}{Jian Wei}, \bibinfo{person}{Jianhua He},
  \bibinfo{person}{Kai Chen}, \bibinfo{person}{Yi Zhou}, {and}
  \bibinfo{person}{Zuoyin Tang}.} \bibinfo{year}{2017}\natexlab{}.
\newblock \showarticletitle{Collaborative filtering and deep learning based
  recommendation system for cold start items}.
\newblock \bibinfo{journal}{\emph{Expert Systems with Applications}}
  \bibinfo{volume}{69} (\bibinfo{year}{2017}), \bibinfo{pages}{29--39}.
\newblock


\bibitem[\protect\citeauthoryear{Wu and Goodman}{Wu and Goodman}{2018}]%
        {wu2018multimodal}
\bibfield{author}{\bibinfo{person}{Mike Wu} {and} \bibinfo{person}{Noah
  Goodman}.} \bibinfo{year}{2018}\natexlab{}.
\newblock \showarticletitle{Multimodal generative models for scalable
  weakly-supervised learning}. In \bibinfo{booktitle}{\emph{Advances in Neural
  Information Processing Systems}}. \bibinfo{pages}{5575--5585}.
\newblock


\bibitem[\protect\citeauthoryear{Xiong, Merity, and Socher}{Xiong
  et~al\mbox{.}}{2016}]%
        {xiong2016dynamic}
\bibfield{author}{\bibinfo{person}{Caiming Xiong}, \bibinfo{person}{Stephen
  Merity}, {and} \bibinfo{person}{Richard Socher}.}
  \bibinfo{year}{2016}\natexlab{}.
\newblock \showarticletitle{Dynamic memory networks for visual and textual
  question answering}. In \bibinfo{booktitle}{\emph{International conference on
  machine learning}}. \bibinfo{pages}{2397--2406}.
\newblock


\bibitem[\protect\citeauthoryear{Yao, Sheng, Wang, Zhang, and Qin}{Yao
  et~al\mbox{.}}{2018}]%
        {yao2018collaborative}
\bibfield{author}{\bibinfo{person}{Lina Yao}, \bibinfo{person}{Quan~Z Sheng},
  \bibinfo{person}{Xianzhi Wang}, \bibinfo{person}{Wei~Emma Zhang}, {and}
  \bibinfo{person}{Yongrui Qin}.} \bibinfo{year}{2018}\natexlab{}.
\newblock \showarticletitle{Collaborative location recommendation by
  integrating multi-dimensional contextual information}.
\newblock \bibinfo{journal}{\emph{ACM Transactions on Internet Technology
  (TOIT)}} \bibinfo{volume}{18}, \bibinfo{number}{3} (\bibinfo{year}{2018}),
  \bibinfo{pages}{1--24}.
\newblock


\bibitem[\protect\citeauthoryear{Yuan, Shalaby, Korayem, Lin, AlJadda, and
  Luo}{Yuan et~al\mbox{.}}{2016}]%
        {yuan2016solving}
\bibfield{author}{\bibinfo{person}{Jianbo Yuan}, \bibinfo{person}{Walid
  Shalaby}, \bibinfo{person}{Mohammed Korayem}, \bibinfo{person}{David Lin},
  \bibinfo{person}{Khalifeh AlJadda}, {and} \bibinfo{person}{Jiebo Luo}.}
  \bibinfo{year}{2016}\natexlab{}.
\newblock \showarticletitle{Solving cold-start problem in large-scale
  recommendation engines: A deep learning approach}. In
  \bibinfo{booktitle}{\emph{2016 IEEE International Conference on Big Data (Big
  Data)}}. IEEE, \bibinfo{pages}{1901--1910}.
\newblock


\bibitem[\protect\citeauthoryear{Zhang, Yao, Sun, and Tay}{Zhang
  et~al\mbox{.}}{2019}]%
        {zhang2019deep}
\bibfield{author}{\bibinfo{person}{Shuai Zhang}, \bibinfo{person}{Lina Yao},
  \bibinfo{person}{Aixin Sun}, {and} \bibinfo{person}{Yi Tay}.}
  \bibinfo{year}{2019}\natexlab{}.
\newblock \showarticletitle{Deep learning based recommender system: A survey
  and new perspectives}.
\newblock \bibinfo{journal}{\emph{ACM Computing Surveys (CSUR)}}
  \bibinfo{volume}{52}, \bibinfo{number}{1} (\bibinfo{year}{2019}),
  \bibinfo{pages}{5}.
\newblock


\bibitem[\protect\citeauthoryear{Zhao, Wang, Dong, and Tian}{Zhao
  et~al\mbox{.}}{2019}]%
        {zhao2019learning}
\bibfield{author}{\bibinfo{person}{Liang Zhao}, \bibinfo{person}{Yang Wang},
  \bibinfo{person}{Daxiang Dong}, {and} \bibinfo{person}{Hao Tian}.}
  \bibinfo{year}{2019}\natexlab{}.
\newblock \showarticletitle{Learning to Recommend via Meta Parameter
  Partition}.
\newblock \bibinfo{journal}{\emph{arXiv preprint arXiv:1912.04108}}
  (\bibinfo{year}{2019}).
\newblock


\bibitem[\protect\citeauthoryear{Zhu, Lin, He, Wang, Guan, Liu, and Cai}{Zhu
  et~al\mbox{.}}{2019}]%
        {zhu2019addressing}
\bibfield{author}{\bibinfo{person}{Yu Zhu}, \bibinfo{person}{Jinghao Lin},
  \bibinfo{person}{Shibi He}, \bibinfo{person}{Beidou Wang},
  \bibinfo{person}{Ziyu Guan}, \bibinfo{person}{Haifeng Liu}, {and}
  \bibinfo{person}{Deng Cai}.} \bibinfo{year}{2019}\natexlab{}.
\newblock \showarticletitle{Addressing the item cold-start problem by
  attribute-driven active learning}.
\newblock \bibinfo{journal}{\emph{IEEE Transactions on Knowledge and Data
  Engineering}} (\bibinfo{year}{2019}).
\newblock


\end{thebibliography}


\clearpage
\appendix
\section{Appendix}
\label{sec:appendix}
\subsection{Dataset Details}
\label{sec:dataset_details}
\subsubsection{Movielens.}
The user profile includes gender (male or female), age group (under 18, 18-25, 26-35, 36-45, 46-50, 51-56, 56+), and 21 different occupations (e.g. student, engineer). The item profile includes movie's release year, genres (e.g. action or adventure), director (one or more directors), and rate (e.g. R, PG).
The mean rating value is 3.58, and the rating comment time is between 2000-04-26 and 2003-03-01. In the raw data, each user has minimum 20 rating histories. We sort the items rated by a user according to their review time, and we trim the dataset, i.e. each user having 20 rating records, to force the model to learn from few shot cases.
\subsubsection{Book-crossing}
The raw Book-crossing data contains many missing and misleading values. The user information includes user age and location. The original user age ranges from 0 to 237, which is in contrast of real cases. Thus, we control the age interval as 5 to 110. The location includes city, state, and country. We only keep the country information for both missing value and privacy consideration. The filtered data has users from 65 different countries. The item information includes the publication year (ranges from 1500 to 2010), author (25593 unique authors), and publisher (5254 unique publishers). 
The Book-crossing dataset does not contain time information for the ratings, thus we assume the data stored in the public dataset are based on the review time, where we keep the order of items that rated by a user to make our dataset. 
We also keep 20 rating records for each user for few-shot learning.

\subsection{Compared methods}
\label{sec:appendix_comparison}
\subsubsection{Code.} 
The code of MeLU\footnote{\url{https://github.com/hoyeoplee/MeLU}} and $S^{2}$ Meta\footnote{\url{https://github.com/THUDM/ScenarioMeta}} are provided by the authors. We modify the input and evaluation modules to fit our experimental settings. For MetaCS-DNN, which has similar idea of MeLU, we modify the code of MeLU to implement it. As for RUM, the code is not published; thus, we implement them with Pytorch. 
\subsubsection{Parameter configuration.}
For MeLU, we use the default parameter settings in the published code. For $S^{2}$ Meta, we implement the code with the parameter settings for Movielens Dataset. For other two comparison methods, we use the suggested parameters when reproduce them. For MetaCS-DNN, the global update learning rate is set to 0.4. For RUM, the learning rate of SGD is determined by grid search in the range of [1, 0.1, 0.01, 0.001, 0.0001], and the number of memory slot K is empirically set as 20. The MERGE parameter $\alpha$ is searched in the range of [0,1] with step 0.1. The embedding dimension and regularization parameters are determined by grid search in the range of [10, 20, 30, 40, 50] and [0.1, 0.01, 0.001, 0.0001], respectively.
\subsubsection{The evaluation metrics in cold-scenarios.}
We provided our definitions for cold-users and cold-items in section \ref{sec:cold_start} and evaluated the performance under the evaluation metrics $MAE$ (see Eq.~\ref{eq:mae_def}) and $NDCG@N$ (see Eq.~\ref{eq:ndcg_def1}). Notice that the users could be either warm users or cold users, and the items rated by a user could be either warm items or cold items -- we label each rating record as in four cold-start scenarios. For example, for a cold user, the rated items are labeled as either warm items (C-W) or cold items (C-C). 
The calculation for $MAE$ in four cold-start scenarios is easy, where we can simply calculate the mean value for the ratings in the four scenarios (i.e. W-W, W-C, C-W, and C-C). 
While for $NDCG@N$, the query set for a user may contain less than $N$ records in different scenarios. Thus, for each scenario, we concatenate the results; separate the results into small clips; calculate $NDCG@N$ for each clip; and then take the mean value of the clips as the results. 

\subsection{Parameter settings.}
Our code is implemented with PyTorch~\footnote{\url{https://pytorch.org/}} 1.4.0 in Python 3.7 and runs on a Linux server with NVIDIA TITAN X. The processed datasets will take about 2GB hard disk space.
The default activation function is LeakyRelu~\footnote{\url{https://en.wikipedia.org/wiki/Rectifier_(neural_networks)}}. 
For Movielens dataset: the dimension of the embedding is set to 100; the default setting of the number of layers is 2; the local learning rate $\rho$ is 0.01 and the global learning rate $\lambda$ is 0.05; the hyper-parameters $\alpha$, $\beta$, $\gamma$ and $\tau$ are set to 0.5, 0.05, 0.1, and 0.1, respectively.
For Bookcrossing dataset: the dimension of the embedding is set to 50; the default setting of the number of layers is 2; the local learning rate $\rho$ is 0.01 and the global learning rate $\lambda$ is 0.01; the hyper-parameters $\alpha$, $\beta$, $\gamma$ and $\tau$ are set to 0.5, 0.1, 0.1, and 0.15, respectively. 
The random seed may affect the results -- sometimes the results may show better or worse performance than our presented results.
The running time for one epoch over all users is about half an hour; so, a strategy is updating the global parameters after learning from a batch of training users. 
During the test process, the model updates as in algorithm~\ref{alg:testing_process}.

\begin{algorithm}[h]
\caption{Testing process of MAMO}
\label{alg:testing_process}
\LinesNumbered
\KwIn{Testing user set $U^{test}$; User profile $\{p_{u}|u\in U^{test}\}$; Item profile $\{p_{i}|i\in (I^{S}_{u}, I^{Q}_{u})\}$; User ratings $\{y_{u,i}|u\in U^{test}, i\in (I^{S}_{u},I^{Q}_{u}) \}$; Hyper-parameters $\alpha$, $\beta$, $\gamma$, $\tau$, $\rho$, $\lambda$; Meta parameters $\phi_{u}$, $\phi_{i}$, $\phi_{r}$, $M_{P}$, $M_{U}$, $M_{U,I}$}
\KwOut{Predicted user preference $\{y_{u,i}|u\in U^{test}, i\in I^{Q}_{u} \}$ }
\For{$u\in U^{test}$}{
  Calculate bias term $b_{u} \leftarrow (p_{u}, M_{U}, M_{P})$ by Eq.~(\ref{eq:mu_attention}-\ref{eq:mu_bu})\;
  Initialize the local parameters $\theta_{i}$, $\theta_{u}$, $\theta_{r}$ by Eq.~(\ref{eq:local_initialize1}-\ref{eq:local_initialize2})\;
  Initialize the preference memory $M_{u,I}\leftarrow (p_{u}, M_{U,I})$ by Eq.~(\ref{eq:mui})\;
  \For{$i \in I^{S}_{u}$}{
    Get user and item embedding $e_{u}$ and $e_{i}$ by Eq.~(\ref{eq:embedding})\;
    Get prediction of $\hat{y_{u,i}}$ by Eq.~(\ref{eq:rec})\;
    Local update $M_{u,I}$\;
    Local update $\theta_{u}$, $\theta_{i}$, $\theta_{r}$ by: $\theta_{*} \leftarrow \theta_{*} - \rho \cdot \nabla_{\theta_{*}}\mathcal{L}(y_{u,i}, \hat{y_{u,i}})$\;
  }
  \For{$i \in I^{Q}_{u}$}{
    Get user and item embedding $e_{u}$ and $e_{i}$ by Eq.~(\ref{eq:embedding})\;
    Get prediction of $\hat{y_{u,i}}$ by Eq.~(\ref{eq:rec})\;
  }
}
\end{algorithm}


\end{document}